\documentclass[aps,prd,floatfix]{revtex4} 

\usepackage{graphics}
\usepackage{graphicx}

\usepackage{latexsym}
\usepackage[cp866]{inputenc}
\usepackage[english]{babel}

\input epsf

\begin{document}

\title{\boldmath Semileptonic decays $D\to\eta\pi e^+\nu_e$
in the $a_0(980)$ region}

\author{N.~N.~Achasov,$^{1}$\footnote{achasov@math.nsc.ru}
A.~V.~Kiselev,$^{1,2}$\footnote{kiselev@math.nsc.ru} and
G.~N.~Shestakov\,$^{1}$\footnote{shestako@math.nsc.ru}}
\affiliation{$^1$\,Laboratory of Theoretical Physics, S.~L.~Sobolev
Institute for Mathematics, 630090 Novosibirsk, Russia,
\\$^2$\,Novosibirsk State University, 630090 Novosibirsk, Russia}


\begin{abstract}
The mechanism of the four-quark production of the light scalar
isovector four-quark state $a_0(980)$ in the $D\to\eta\pi e^+\nu_e$
decays is discussed. It is shown that the characteristic features of
the shape of the $\eta\pi$ mass spectra expected in our scheme can
serve as the indicator of the production mechanism and internal
structure of the $a_0(980)$ resonance.
\end{abstract}

\maketitle

\section{INTRODUCTION}

Semileptonic decays of $D$ and $B$ mesons into $e^+\nu_e$ and a pair
of pseudoscalar mesons are a good laboratory for studying the quark
structure of light scalar mesons \cite{CLEO09b,WL10,Pe10,Fa11,AK12,
PDG2020}. The first important experiments in this direction were
carried out by the Collaborations {\it BABAR}, on the decay
$D^+_s\to f_0(980)e^+\nu_e\to K^+K^-e^+\nu_e$ \cite{BABAR08}, CLEO,
on the decay $D^+_s\to f_0(980)e^+\nu_e\to\pi^+ \pi^-e^+\nu_e$
\cite{CLEO09a, CLEO09b}, and BESIII, on the decays $D^0\to
a_0(980)^-e^+\nu_e\to\eta\pi^-e^+\nu_e$ \cite{BESIII18}, $D^+\to
a_0(980)^0e^+\nu_e\to\eta\pi^0e^+\nu_e$ \cite{BESIII18}, $D^+\to
f_0(500)e^+\nu_e\to\pi^+\pi^-e^+\nu_e$ \cite{BESIII19}, and
$D^+_s\to a_0(980)^0e^+\nu_e\to\pi^0\eta e^+\nu_e$ \cite{BESIII21}.
Theoretical studies of these decays and phenomenological data
processing have been carried out especially recently very
intensively \cite{WL10,Pe10,Fa11,AK12,AK14,SO15,Sh17,Ch17,AK18,AK19,
Ac20,So20, AKS20,Sh21a,Sh21b}. In many ways, their goal is to find
new evidence in favor of the supposed exotic four-quark ($q^2\bar
q^2$) structure of the $f_0(500)$, $f_0(980)$, and $a_0(980)$
resonances or indications that contradict this. They also contain
the descriptions of the mass distributions of pseudoscalar meson
pairs and calculations of the relative and absolute values of the
probabilities of the indicated semileptonic decays of $D$ and $B$
mesons, taking into account the requirements of the unitarity
condition.

We continue the study of the decays $D^0\to d\bar ue^+\nu_e \to
a_0(980)^-e^+\nu_e\to\eta\pi^- e^+\nu_e$ and $D^+\to d\bar de^+
\nu_e\to a_0(980)^0e^+\nu_e\to\eta \pi^0e^+\nu_e$ started in Ref.
\cite{AK18}. These decays are interesting in that they provide
direct probing of constituent two-quark $d\bar u$ and $(u\bar
u-d\bar d)/\sqrt{2}$ components in the wave functions of the
$a_0(980)^-$ and $a_0(980)^0$ mesons, respectively \cite{AK12}. In
Ref. \cite{AK18} the mass spectra of $\eta\pi$ pairs were presented
for the case when the $a_0(980)$ states have no two-quark components
at all. It was assumed that the production of the four-quark
$a_0(980)$ state occurs via its mixing with the heavy $q\bar q$
state $a'_0(1400)$, due to their common decay channels into
$\eta\pi$, $\eta'\pi$, and $K\bar K$. Here we also assume that the
$a_0(980)$ meson has a four-quark structure \cite{Ja77,AI89}. The
difference is that we consider the mechanism of four-quark
fluctuations of $d\bar u$ and $d\bar d$ sources, $d\bar u\to d\bar
qq\bar u$ and $d\bar d\to d\bar qq\bar d$, which in the language of
two-body hadronic states means the creation of the $\eta\pi$,
$\eta'\pi$, and $K\bar K$ pairs, which are then dressed by strong
interactions in the final state. We elucidate the characteristic
features of this mechanism and do not involve in consideration a
heavy state of the $a'_0(1400)$ type. A similar approach was used in
Refs. \cite{AKS20,Ac21} to describe the CLEO \cite{CLEO09b} and
BESIII \cite{BESIII19} data on the decays $D^+_s\to f_0(980)e^+\nu_e
\to\pi^+\pi^-e^+\nu_e$ and $D^+\to f_0(500)e^+\nu_e \to\pi^+\pi^-
e^+\nu_e$ as well as the BESIII \cite{Ab15} data on the $J/\psi\to
\gamma\pi^0\pi^0$ decay in the $\pi^0\pi^0$ invariant mass region
from the threshold up to 1 GeV.

This paper is organized as follows. In Sec. II after a brief
discussion of the experimental situation, the general formulas are
given for the widths of semileptonic decays $D^0\to\eta\pi^-e^+
\nu_e$ and $D^+\to \eta\pi^0e^+\nu_e$ with the production of the
$\eta\pi$ system in the $S$ wave. Section III is devoted to a
discussion of the four-quark mechanism of the $a_0(980)$ resonance
production in the $D\to\eta\pi e^+\nu_e$ decays. We show that the
characteristic features of the shape of the $\eta\pi^-$ and $\eta
\pi^0$ mass spectra expected in our scheme can serve as an indicator
of the production mechanism and internal structure of the $a_0(980)$
state. In Sec. IV we discuss the decays $D^0\to K^0K^-e^+\nu_e$,
$D^+\to K^0\bar K^0e^+\nu_e$, and $D^+\to K^+K^-e^+ \nu_e$ which
also are of interest in connection with the production of light
scalar mesons. However, these decays are strongly suppressed by the
phase space near the $K\bar K$ thresholds and their experimental
investigations are little realistic at present. In Sec. V the
conclusions are briefly formulated.


\section{BESIII data and decay widths}

Recently, the BESIII  Collaboration \cite{BESIII18} obtained the
first data on semileptonic decays $D^0\to a_0(980)^-e^+ \nu_e$ and
$D^+\to a_0(980)^0e^+\nu_e$ using the reaction $e^+e^-\to\psi(3770)
\to D\bar D$ at a center-of-mass energy of 3.773 GeV and the tagged
$D$ meson technique \cite{Ba86}. They selected $25.7^{+6.4}_{-5.7}$
signal events for $D^0 \to a_0(980)^-e^+\nu_e$ and $10.2^{+ 5.0}_{
-4.1}$ signal events for $D^+\to a_0(980)^0e^+\nu_e$. The
statistical significance of signal is $6.4\sigma$ for $D^0 \to
a_0(980)^-e^+\nu_e$ and $2.9\sigma$ for $D^+\to a_0(980)^0e^+\nu_e$.
The absolute branching fractions and the ratio of decay widths were
determined to be \cite{BESIII18}
\begin{eqnarray}\label{Eq1}
\mathcal{B}(D^0\to a_0(980)^-e^+\nu_e)\times\mathcal{B}(a_0(980)^-
\to\eta\pi^-)&=&[1.33^{+0.33}_{-0.29}(\mbox{stat})\pm0.09(\mbox{syst})]
\times10^{-4},\\ \label{Eq2} \mathcal{B}(D^+\to a_0(980)^0e^+\nu_e)
\times\mathcal{B} (a_0(980)^0 \to\eta\pi^0)&=&[1.66^{+0.81}_{-0.66}
(\mbox{stat})\pm0.11(\mbox{syst})] \times10^{-4},\\ \label{Eq3}
\frac{\Gamma(D^0\to a_0(980 )^-e^+ \nu_e)}{\Gamma(D^+\to a_0(980)^0
e^+\nu_e)} &=&2.03\pm0.95\pm0.06.\end{eqnarray} When obtaining Eq.
(\ref{Eq3}), it was assumed that $\mathcal{B}(a_0(980)^-\to\eta
\pi^-)=\mathcal{B}(a_0(980)^0\to\eta\pi^0)$.

\begin{figure} [!ht] 
\begin{center}\includegraphics[width=13cm]{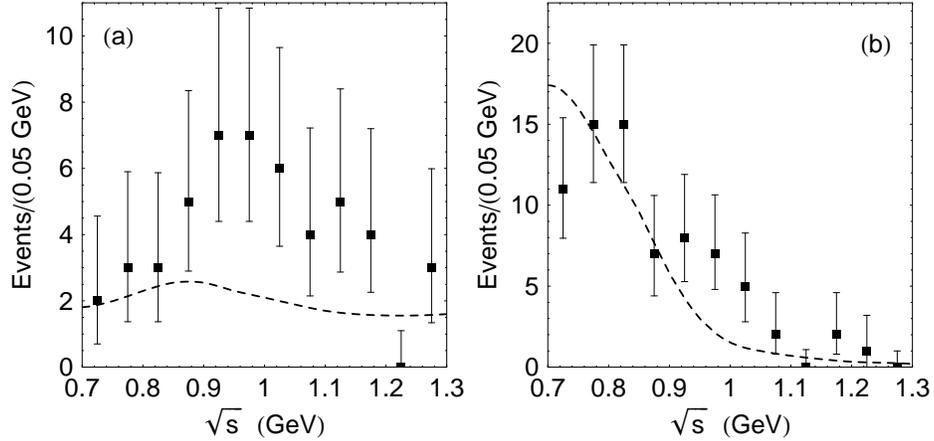}
\caption{\label{Fig1} The $\eta\pi$ mass spectra (a) for $D^0\to
(a_0(980)^-\to\eta\pi^-)e^+\nu_e$ and (b) for $D^+\to(a_0(980)^0
\to\eta\pi^0)e^+\nu_e$. Points with error bars are the BESIII data
\cite{BESIII18}. The dashed curves show the sum of the background
contributions defined by BESIII.}
\end{center}\end{figure}

In the measured $\eta\pi$ mass spectra in the range of the invariant
mass of the $\eta\pi$ system, $m_{\eta\pi}\equiv\sqrt{s}$, from
$0.7$ GeV to $1.3$ GeV, there is a complex configuration of
background contributions \cite{BESIII18} (see Fig. \ref{Fig1}).
A small number of signal events, a noticeable background and a wide
step in $\sqrt{s}$ (equal to $50$ MeV) do not allow us to clearly
see in the $\eta\pi^-$ and $\eta\pi^0$ mass spectra the line shape
of the $a_0(980)$ resonance. It is clear that the high statistics on
the decays $D\to a_0(980)e^+ \nu_e$ is highly demanded by the
physics of light scalar mesons.

\begin{figure} [!ht] 
\begin{center}\includegraphics[width=6.7cm]{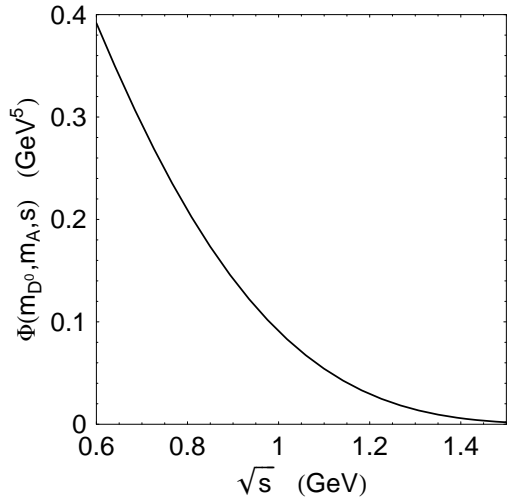}
\caption{\label{Fig2} The function $\Phi(m_{D^0},m_A,s)$ at
$m_A=m_{D^+_{1}}=2.42$ GeV.}\end{center}\end{figure}

Let us write the differential width for the $D^0=c\bar u$ decay into
$\eta\pi^-e^+\nu_e$ in the form (see, for example, Refs.
\cite{CLEO09b,BESIII19,Sa11})
\begin{eqnarray}\label{Eq4}\frac{d^2\Gamma_{D^0\to d\bar ue^+
\nu_e\to(S^-\to\eta\pi^-)e^+ \nu_e}(s,q^2)}{d\sqrt{s}\,dq^2}=
\frac{G^2_F|V_{cd}|^2}{24\pi^3} p^3_{\eta\pi^-}(m_{D^0},q^2,
s)|f^{D^0}_+(q^2)|^2 \frac{2\sqrt{s}}{\pi}|F^{D^0}_{d\bar u\to
S^-\to\eta \pi^-}(s)|^2\rho_{\eta\pi^-}(s),\end{eqnarray} where $s$
and $q^2$ are the invariant mass squared of the virtual scalar state
$S^-$ (or the $S$-wave $\eta\pi^-$ system) and the $e^+\nu_e$
system, respectively; $G_F$ is the Fermi constant,
$|V_{cd}|=0.221\pm0.004$ is a Cabibbo-Kobayshi-Maskawa matrix
element \cite{PDG2020}; $p_{\eta\pi^-}$ is the magnitude of the
three-momentum of the $\eta\pi^-$ system in the $D^0$ meson rest
frame,
\begin{eqnarray}\label{Eq5}
p_{\eta\pi^-}(m_{D^0},q^2,s)=\sqrt{[(m_{D^0}-\sqrt{s})^2-q^2]
[(m_{D^0}+ \sqrt{s})^2-q^2]}/(2m_{D^0}),\end{eqnarray} and
$\rho_{\eta\pi^-}(s)=\sqrt{[s-(m_\eta+m_{\pi^-})^2]
[s-(m_\eta-m_{\pi^-})^2]}/s$. In a simplest pole approximation, the
form factor $f^{D^0}_+(q^2)$ has the form \cite{CLEO09b,BESIII19,
Sa11}
\begin{eqnarray}\label{Eq6} f^{D^0}_+(q^2)=\frac{f^{D^0}_+(0)}{1-
q^2/m^2_A},\end{eqnarray} where we put $m_A=m_{D^+_{1}}=2.42$ GeV
\cite{PDG2020} (in principle, $m_A$ can be extracted from the data
by fitting). The amplitude $F^{D^0}_{d\bar u\to S^-\to
\eta\pi^-}(s)$ describes the formation and decay into $\eta\pi^-$ of
the virtual scalar isovector state $S^-$ produced in the
$D^0\to\eta\pi^-e^+\nu_e$ decay. The $\eta\pi^-$ invariant mass
distribution integrated over the full $q^2$ region is given by
\begin{eqnarray}\label{Eq7}
\frac{d\Gamma_{D^0\to d\bar ue^+\nu_e\to(S^-\to\eta\pi^-)e^+
\nu_e}(s)}{d\sqrt{s}}=\frac{G^2_F|V_{cq}|^2}{24\pi^3}|f^{D^0}_+(0)
|^2\,\Phi(m_{D^0},m_A,s)\frac{2\sqrt{s}}{\pi}|F^{D^0}_{d\bar u\to
S^-\to\eta\pi^-}(s)|^2\rho_{\eta\pi^-}(s),\end{eqnarray} where the
function $\Phi(m_{D^0},m_A,s)$ is
\begin{eqnarray}\label{Eq8}
\Phi(m_{D^0},m_A,s)=\int\limits_0^{(m_{D^0}-\sqrt{s})^2}\frac{p^3_{
\eta\pi^-}(m_{D^0},q^2,s)}{|1-q^2/m^2_A|^2}\,dq^2.\end{eqnarray} As
can be seen from Fig. \ref{Fig2}, it notably enhances the
$\eta\pi^-$ mass spectrum as $\sqrt{s}$ decreases.

The differential widths of the semileptonic decay of the $D^+=c\bar
d$ meson into the $S$-wave $\eta\pi^0$ system are described by the
formulas similar to Eqs. (\ref{Eq4})--(\ref{Eq8}).

\section{\boldmath Four-quark production mechanism of the
four-quark $a_0(980)$ resonance}

\begin{figure} [!ht] 
\begin{center}\includegraphics[width=6cm]{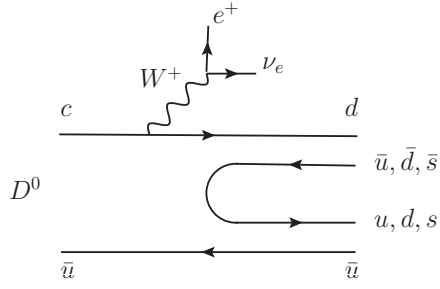}
\caption{\label{Fig3} The OZI allowed four-quark fluctuations of the
$d\bar u$ source, $d\bar u\to d\bar qq\bar u$.}
\end{center}\end{figure}

We start with the semileptonic decay of the $D^0=c\bar u$ meson
$D^0\to d\bar ue^+\nu_e\to(S^-\to\eta\pi^-)e^+ \nu_e$. Let, as a
result of radiation of the lepton pair $e^+\nu_e$ by the valence $c$
quark included in $D^0$, a virtual system of $d$ and $\bar u$ quarks
in the scalar state is produced. Consider the four-quark
fluctuations of such a $d\bar u$ source, $d\bar u\to d\bar qq\bar
u$, corresponding to the diagram shown in Fig. \ref{Fig3}. The
limitation to this type of fluctuations is related to the
Okubo-Zweiga-Iizuka (OZI) rule \cite{Ok63,Ok78,Zw80,Ii66}, according
to which other fluctuations such as annihilation or creation of
$q\bar q$ pairs, corresponding to the so-called ``hair-pin''
diagrams, are suppressed.
In the language of hadronic states the seed four-quark fluctuations
$d\bar u\to d\bar qq\bar u$ imply the production of $\eta\pi^-$,
$\eta'\pi^-$, and $K^0K^-$ meson pairs. As a first approximation, we
assume that the amplitude of the $d\bar u\to d\bar qq\bar u$
transition, which we denote by $g_0$, does not depend on the flavor
of the light $q$ quark. Then for the hadron production constants
$g_{d\bar u\eta\pi^-}$, $g_{d\bar u\eta'\pi^-}$, and $g_{d \bar
uK^0K^-}$ the following relations hold:
\begin{eqnarray}\label{Eq10} g_{d\bar
u\eta\pi^-}=\sqrt{2}g_0\sin(\theta_i-\theta_p),\ \ \ g_{d\bar
u\eta'\pi^-}=\sqrt{2}g_0\cos(\theta_i-\theta_p),\ \ \ g_{d\bar uK^0
K^-}=g_0.\end{eqnarray} Here $\theta_i=35.3^\circ$ is the so-called
``ideal'' mixing angle and $\theta_p=-11.3^\circ$ is the mixing
angle in the nonet of the light pseudoscalar mesons \cite{PDG2020}.
The first two relations in Eq. (\ref{Eq10}) are obtained taking into
account the expansion of the state containing non-strange quarks
$\eta_n=(u\bar u+d\bar d)/\sqrt{2}$, which is created in a pair with
$\pi^-$, in terms of physical states $\eta$ and $\eta'$ with
definite masses: $\eta_n=\eta\sin(\theta_i-\theta_p)+\eta'\cos
(\theta_i-\theta_p)$.

\begin{figure} [!ht] 
\begin{center}\includegraphics[width=14cm]{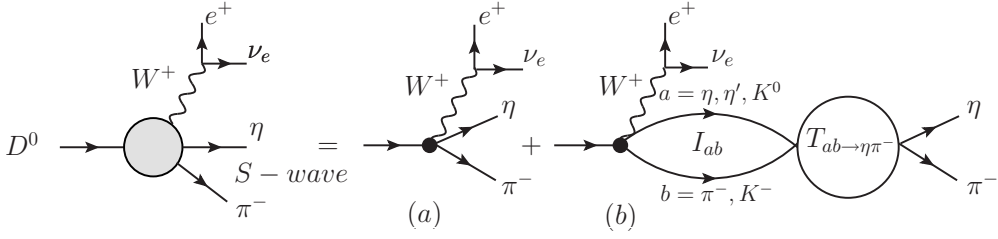}
\caption{\label{Fig4} The production mechanism of the four-quark
$a_0(980)^-$ resonance in the decay $D^0\to\eta\pi^-e^+
\nu_e$.}\end{center}\end{figure}

\begin{figure} [!ht] 
\begin{center}\includegraphics[width=14cm]{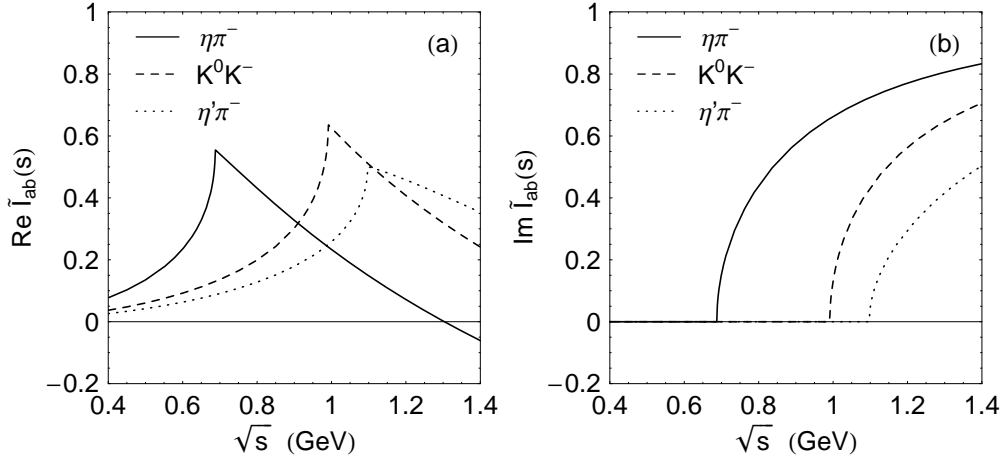}
\caption{\label{Fig5} Line shapes of (a) $\mbox{Re}\widetilde{
I}_{ab}(s)$ and (b) $\mbox{Im}\widetilde{I}_{ab}(s)$ for the
intermediate states $ab=\eta\pi^-$, $K^0K^-$, and $\eta'\pi^-$.}
\end{center}\end{figure}

Thus, the production of the four-quark $a_0(980)^-$ resonance can
occur as a result of the fact that the seed four-quark fluctuations
$d\bar u\to\eta \pi^-,\,\eta'\pi^-,\,K^0K^-$ are dressed by strong
interactions in the final state. The described picture of the
$a_0(980)^-$ production in the language of hadronic diagrams is
shown in Fig. \ref{Fig4}.
According to this figure, we write the amplitude $F^{D^0}_{d\bar
u\to S^-\to\eta\pi^-}(s)$ from Eq. (\ref{Eq7}) in the form
\begin{eqnarray}\label{Eq11}
F^{D^0}_{d\bar u\to S^-\to\eta\pi^-}(s)=g_{d\bar u\eta\pi^-}\left[1+
I_{\eta\pi^-}(s)\, T_{\eta\pi^-\to\eta\pi^-}(s)\right]+g_{d\bar
u\eta'\pi^-}I_{\eta'\pi^-}(s)\,T_{\eta'\pi^-\to\eta\pi^-}(s)
\nonumber \\+ g_{d\bar uK^0 K^-}I_{K^0 K^-}(s)\,T_{K^0K^-\to\eta
\pi^-}(s),\end{eqnarray} where $I_{ab}(s)$ is the amplitude of the
two-point loop diagram with $ab$ intermediate state and $T_{ab\to
\eta\pi^-}(s)$ is the amplitude of the $S$-wave $ab\to\eta\pi^-$
transition; $ab=\eta\pi^-,\,\eta' \pi^-,\,K^0K^-$. The loop
amplitude $I_{ab}(s)$ has the form
\begin{eqnarray}\label{Eq12} I_{ab}(s)=C_{ab}+\widetilde{I}_{ab}(s)=
C_{ab}+\frac{s}{\pi}\int\limits^\infty_{m_{ab}^{(+)\,2}}\frac{
\rho_{ab}(s')\,ds'}{ \,s'(s'-s-i\varepsilon)}\,,\end{eqnarray} where
the function $\widetilde{I}_{ab}(s)$ is the singly subtracted at
$s=0$ dispersion integral. Explicit expressions for
$\widetilde{I}_{ab}(s)$ in different regions of $s$ are given in the
Appendix. The real and imaginary parts of this function for
different $ab$ are shown in Fig. \ref{Fig5}. The quantity $C_{ab}$
in Eq. (\ref {Eq12}) is a subtraction constant. The choice of these
constants will be discussed below.

We saturate the amplitudes $T_{ab\to\eta\pi^-}(s)$ with the
contribution of the four-quark $a_0(980)^-$ resonance. The flavor
structure of its wave function has the form \cite{Ja77}
\begin{eqnarray}\label{Eq13} a_0(980)^-=\bar u\bar sds=\frac{1}{
\sqrt{2}}\,\eta_s\pi^--\frac{1}{\sqrt{2}}\,K^0K^-.\end{eqnarray}
Considering that $\eta_s=s\bar s=\eta'\sin(\theta_i-\theta_p)
-\eta\cos(\theta_i-\theta_p)$, for the coupling constants of the
$a_0(980)^-$ with the superallowed decay channels into pairs of
pseudoscalar mesons, the following relations hold:
\begin{eqnarray}\label{Eq14} g_{a^-_0\eta\pi^-}=\bar{g}\cos(\theta_i-
\theta_p),\ \ \ g_{a^-_0\eta'\pi^-}=-\bar{g}\sin(\theta_i-
\theta_p),\ \ \ g_{a^-_0K^0 K^-}=\bar{g},\end{eqnarray} where
$\bar{g}$ is the overall coupling constant. Thus,
\begin{eqnarray}\label{Eq15} T_{ab\to\eta\pi^-}(s)=\frac{g_{a^-_0ab}
g_{a^-_0\eta\pi^-}}{16\pi}\frac{1}{D_{a^-_0}(s)}\,,\end{eqnarray}
where $D_{a^-_0}(s)$ is the inverse propagator of the $a_0(980)^-$
resonance. It has the form
\begin{eqnarray}\label{Eq16} D_{a^-_0}(s)=
m^2_{a^-_0}-s+\sum_{ab}[\mbox{Re}\Pi^{ab}_{a^-_0}(m^2_{a^-_0})-
\Pi^{ab}_{a^-_0}(s)],\end{eqnarray} where $m_{a^-_0}$ is the
$a_0(980)^-$ mass and $\Pi^{ab}_{a^-_0}(s)$ stands for the matrix
element of the $a_0(980)^-$ polarization operator corresponding to
the contribution of the $ab$ intermediate state. Re$\Pi^{ab}_{a^-_0}
(s)$ is defined by the singly subtracted at $s=0$ dispersion
integral of
\begin{eqnarray}\label{Eq17}
\mbox{Im}\,\Pi^{ab}_{a^-_0}(s)=\sqrt{s}\Gamma_{a^-_0\to ab}(s)=
\frac{g^2_{a^-_0 ab}}{16\pi}\rho_{ab}(s)\,\theta(\sqrt{s}-m_a-m_b),
\end{eqnarray} i.e., $\Pi^{ab}_{a^-_0}(s)=g^2_{a^-_0
ab}\,\widetilde{I}_{ab}(s)/(16\pi)$. The terms $\sum_{ab}[\mbox{Re}
\Pi^{ab}_{a^-_0}(m^2_{a^-_0})- \Pi^{ab}_{a^-_0}(s)]$ in Eq.
(\ref{Eq16}) take into account the finite width corrections and
guarantee that the Breit-Wigner mass of the $a_0(980)^-$ resonance
is defined by the condition Re$D_{a^-_0}(s=m^2_{a^-_0})=0$.

We now rewrite Eq. (\ref{Eq11}) using Eqs. (\ref{Eq10}),
(\ref{Eq14}), and (\ref{Eq15}) and notation $\vartheta=\theta_i-
\theta_p$ as follows:
\begin{eqnarray}\label{Eq18} F^{D^0}_{d\bar u\to
S^-\to\eta\pi^-}(s)=g_0\sqrt{2}\sin\vartheta\left[1+\left(I_{\eta
\pi^-}(s)-I_{\eta'\pi^-}(s)\right)\frac{\bar{g}^2\cos^2 \vartheta}
{16\pi D_{a^-_0}(s)}\right]+g_0I_{K^0 K^-}(s)\frac{\bar{g}^2 \cos
\vartheta}{16\pi D_{a^-_0}(s)}.\end{eqnarray} This expression shows
that the contributions proportional to the loops $I_{\eta \pi^-}(s)$
and $I_{\eta'\pi^-} (s)$ enter with different signs and therefore
partially cancel each other. This cancellation implements the OZI
rule in the language of hadronic intermediate states \cite{Lip87,
Lip96}. Indeed, in the case under consideration, the sum of the
contributions of the $\eta\pi^-$ and $\eta'\pi^-$ loops is due to
the transition $d\bar u\to\eta_n\pi^-$ and then $\eta_n\pi^-$ into
the $\eta_s\pi^-$ component of the $a_0(980)^-$ wave function, which
is suppressed according to the OZI rule.

Let us discuss the choice of the subtraction constants $C_{ab}$ in
the loops $I_{ab}(s)$, see (\ref{Eq12}). The assumption that all
these constants are equal, $C_{\eta \pi^-}=C_{\eta'\pi^-}=
C_{K^0K^-}$, is simplest and most economical in terms of the number
of free parameters. Note that nothing changes if we put $C_{\eta
\pi^-}=C_{\eta'\pi^-} \neq C_{K^0K^-}$, since if $C_{\eta
\pi^-}=C_{\eta'\pi^-}$, the expression (\ref{Eq18}) does not depend
on their value at all, due to the OZI reduction, and depends only on
the parameter $C_{K^0K^-} $. Below we utilize this choice for
$C_{ab}$. In so doing, the model is still quite flexible.

Thus, there are two parameters characterizing the mechanism of the
$a_0(980)^-$ production in $D^0\to a_0(980)^-e^+\nu_e\to\eta
\pi^-e^+\nu_e$ in our model. They are: the product
$f^{D^0}_+(0)g_0$, it determines the general normalization of the
decay width, and the constant $C_{K^0 K^-}$ which essentially
influences on the shape of the $\eta \pi^-$ mass spectrum. The
actual parameters of the $a_0(980)^-$ resonance are its mass
$m_{a^-_0}$ and coupling constant $\bar{g}$, see (\ref{Eq14}).
\begin{figure} [!ht] 
\begin{center}\includegraphics[width=7cm]{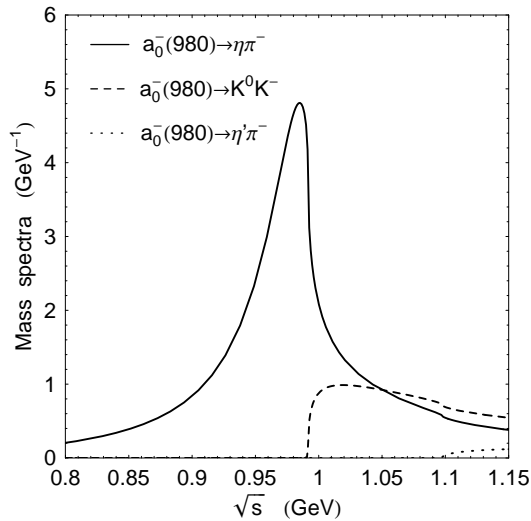}
\caption{\label{Fig6} The Breit-Wigner line shapes of the
$a_0(980)^-$ resonance in $\eta\pi^-$, $K^0K^-$, and $\eta'\pi^-$
decay channels, see Eq. (\ref{Eq19}).}\end{center}\end{figure}
Figure \ref{Fig6} shows as a guide the shapes of the Breit-Wigner
mass distributions
\begin{eqnarray}\label{Eq19} \frac{d\mathcal{B}(a_0(980)^-\to ab;s)}
{d\sqrt{s}}=\frac{2\sqrt{s}}{\pi}\frac{\sqrt{s}\Gamma_{a^-_0\to
ab}(s)}{|D_{a^-_0}(s)|^2}\end{eqnarray} for the solitary
$a_0(980)^-$ resonance in $\eta \pi^-$, $K^0K^-$, and $\eta'\pi^-$
decay channels. They are plotted using Eqs. (\ref{Eq14}),
(\ref{Eq16}), (\ref{Eq17}), and (\ref{Eq19}) at $m_{ a^-_0}=0.985$
GeV and $g^2_{a^-_0\eta\pi^-}/(16\pi)=0.2$ GeV$^2$. Note that the
decay width $\Gamma_{a^-_0\to\eta\pi^-}(s)$ calculated by Eq. (\ref
{Eq17}) at $\sqrt{s}=m_{a^-_0}$ is approximately equal to 132 MeV,
while the visible width of the $a_0(980)^-$ peak at its half-maximum
in $\eta\pi^-$ channel is $\approx55$ MeV (see Fig. \ref {Fig6}).
This narrowing is the consequence of the proximity of $m_{a^-_0}$ to
the $K^0K^-$ threshold and strong coupling of the $a_0(980 )^-$ to
both $\eta\pi^-$ and $K^0K^-$ decay channel \cite{Fl76,ADS80}.

Recently, the BESIII Collaboration observed in the high-statistics
experiment on the decay $\chi_{c1}\to\eta\pi^+\pi^-$ the impressive
peak from $a_0(980)$ resonance in the $\eta\pi$ mass spectra
\cite{BESIII17}. There is a possibility that the $a_0(980)$
resonance will manifest itself in the $D\to\eta\pi e^+\nu_e$ decays
in a similar way. Below we discuss the conditions under which such a
possibility is realized in our model.

\begin{figure} 
\begin{center}\includegraphics[width=7.5cm]{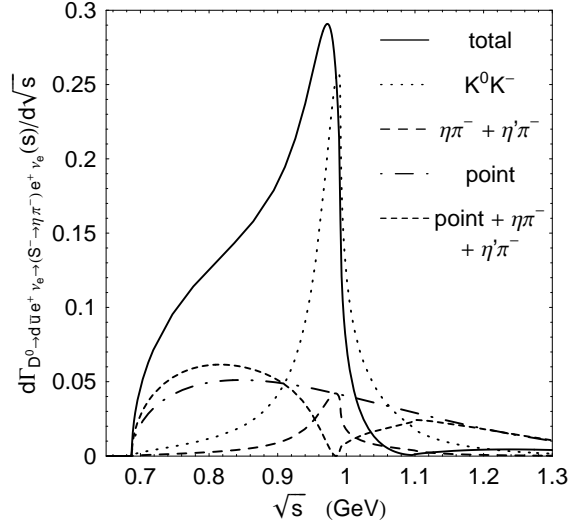}
\caption{\label{Fig7} The solid curve shows an example of the shape
of the $\eta\pi^-$ mass spectrum in the decay $D^0\to d\bar
ue^+\nu_e \to(S^-\to\eta\pi^-)e^+\nu_e$ with a peak in the region of
1 GeV due to the creation of the $a_0(980)^-$ resonance. Detailed
description of the components of this spectrum, shown in the figure
by other curves, is given in the text. The scale along the ordinate
axis is chosen arbitrarily. }\end{center}\end{figure}

The solid curve in Fig. \ref{Fig7} shows an example of the shape of
the $\eta\pi^-$ mass spectrum in the decay $D^0\to d\bar ue^+\nu_e
\to(S^-\to\eta\pi^-)e^+\nu_e$ with a peak in the region of 1 GeV due
to the creation of the $a_0(980)^-$ resonance. The calculation was
done using Eqs. (\ref{Eq7}), (\ref{Eq14}), and (\ref{Eq18}) at the
above values of $m_{a^-_0}$ and $g^2_{a^-_0\eta\pi^-}/(16\pi)$ and
$C_{K^0 K^-}=0.6$. The value of $C_{K^0K^-}$ for this example we
took to be comparable with the value of Re$\widetilde{I}_{K^0
K^-}(s)$ in the region near the $K^0K^-$ threshold, see Fig.
\ref{Fig5}(a). The difference between the solid curves in Figs.
\ref{Fig6} and \ref{Fig7} demonstrates the possible influence of the
production mechanism on the shape of the $a_0(980)^-$ peak in the
$\eta\pi^-$ channel. The dotted curve in Fig. \ref{Fig7} shows the
contribution from the last term in Eq. (\ref{Eq18}) corresponding,
according to diagram (b) in Fig. \ref{Fig4}, to creation of the
$a_0(980)^-$ resonance via the $K^0K^-$ intermediate state. The
dash-dotted curve shows the contribution due to the seed pointlike
diagram (a) in Fig. \ref{Fig4}. In Eq. (\ref{Eq18}), it corresponds
to the term equal to $g_0\sqrt{ 2}\sin\vartheta$. The long dashed
curve shows the total contribution to the $a_0(980 )^-$ production
from the $\eta\pi^-$ and $\eta'\pi^- $ intermediate states (see Fig.
\ref{Fig4}). In Eq. (\ref{Eq18}), this contribution is proportional
to $\left(I_{\eta\pi^-}(s)- I_{\eta'\pi^-}(s) \right)$. Recall that
we consider the case when $C_{\eta\pi^-}=C_{\eta'\pi^-}$. At last,
the short dashed curve in Fig. \ref{Fig7} shows the total
contribution due to the terms in square brackets in Eq.
(\ref{Eq18}). It is seen that these terms strongly compensate each
other in the region $\sqrt{s}\approx m_{a^-_0}$. First, this happens
because in this region Re$\left( \widetilde{I}_{ \eta
\pi^-}(s)-\widetilde{I }_{\eta'\pi^-}(s)\right)\approx0$, as shown
in Fig. \ref{Fig5}(a). Second, the other contributions for $\sqrt{s
}<m_{K^0}+m_{K^- }$ can be represented as $e^{i\delta_{
\eta\pi^-}(s)}\cos\delta_{\eta \pi^-}(s)$, where $\delta_{\eta\pi^-}
(s)$ is the phase of the $S $-wave elastic $\eta\pi^-$ scattering
amplitude. In this case, $\delta_{\eta\pi^-}(s)$ is a purely
resonant phase and therefore $\cos\delta_{ \eta\pi^-}(m^2_{
a^-_0})=0$. Our statement follows from the chain of equalities:
$$1+i\mbox{Im}I_{\eta\pi^-}(s)\frac{\bar{g}^2\cos^2\vartheta}{16\pi
D_{a^-_0}(s)}=1+i\rho_{\eta\pi^-}(s)T_{\eta\pi^-\to\eta\pi^-}(s)=1+
i\rho_{\eta\pi^-}(s)\frac{e^{2i\delta_{\eta\pi^-}(s)}-1}{2\rho_{\eta
\pi^-}(s)}=e^{i\delta_{\eta\pi^-} (s)}\cos\delta_{\eta\pi^-}(s).$$
Thus, the phase of the amplitude $F^{D^0}_{d\bar u\to S^-\to\eta
\pi^- }(s)$, see Eq. (\ref{Eq18}), for $\sqrt{s }<m_{K^0}+ m_{K^-}$
coincides with the phase of the elastic $\eta\pi^-$ scattering in
accordance with the requirement of the unitarity condition
\cite{Wa52}. Equation (\ref{Eq11}), which has a more general form,
also satisfies the unitarity requirement, since in the elastic
region the phases of the amplitudes $T_{\eta'\pi^-\to\eta\pi^-}(s)$
and $T_{K^0K^-\to\eta \pi^-}(s)$ coincide with the phase of the
amplitude $T_{\eta\pi^- \to\eta\pi^-}(s)$ and the functions
$I_{\eta'\pi^-}(s)$ and $I_{K^0 K^-}(s)$ are real. The phenomenon of
compensation of the pointlike production amplitude
[$g_0\sqrt{2}\sin\vartheta$ in Eq. (\ref{Eq18})] by the resonant one
at $\sqrt{s}\approx m_{ \mbox{\scriptsize{res}}}$ is also well
known, see, for example, \cite{Om58,GRS69}.

The relative role of the $K^0K^-$ intermediate state in the $a_0(980
)^-$ production increases with increasing the parameter $C_{K^0K^-}
$. The width of the resonance peak in the $\eta\pi^-$ mass spectrum
narrows, while its height increases and it becomes more pronounced.
However, the characteristic enhancement of the left wing of the
resonance spectrum and a sharp jump of its right wing (see Fig.
\ref{Fig7}), caused by interference between different contributions,
persist when $C_{K^0 K^-}$ changes over a wide range. The specific
form of the $\eta\pi^-$ mass spectrum is directly related to the
considered mechanism of the $a_0(980)^-$ production and therefore
can serve as its indicator.

\begin{figure} [!ht] 
\begin{center}\includegraphics[width=6cm]{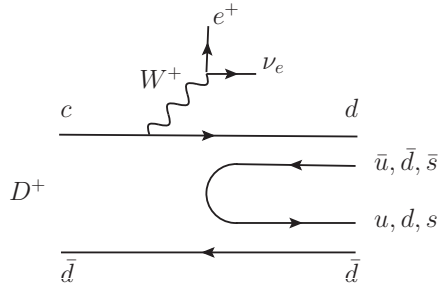}
\caption{\label{Fig8a} The OZI allowed four-quark fluctuations of
the $d\bar d$ source, $d\bar d\to d\bar qq\bar d$.}
\end{center}\end{figure}

\begin{figure} [!ht] 
\begin{center}\includegraphics[width=14cm]{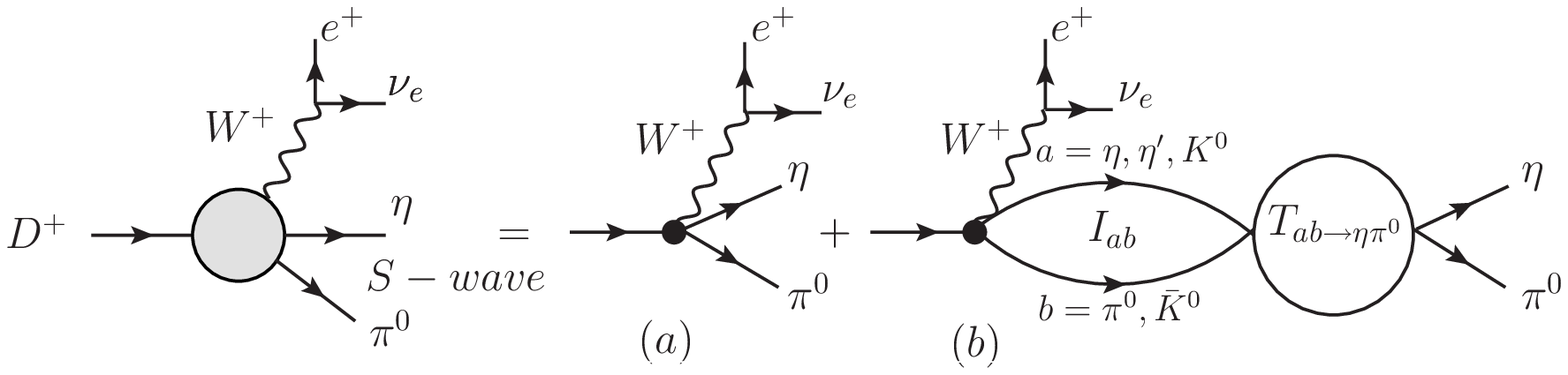}
\caption{\label{Fig8} The production mechanism of the four-quark
$a_0(980)^0$ resonance in the decay $D^+\to\eta\pi^0e^+\nu_e$.
}\end{center}\end{figure}

Decays of $D^+\to a_0(980)^0e^+\nu_e\to\eta\pi^0e^+\nu_e$  and
$D^0\to a_0(980)^-e^+\nu_e\to\eta\pi^- e^+\nu_e$ have the same
mechanisms, as is clear from Figs. \ref{Fig8a}, \ref{Fig8} and
\ref{Fig3}, \ref{Fig4}.
Their amplitudes $F^{D^+}_{d\bar d\to S^0\to\eta \pi^0}(s)$ and
$F^{D^0 }_{d\bar u\to S^-\to\eta \pi^-}(s)$ [see Eq. (\ref{Eq18})]
are related to each other (within a constant phase factor which can
be chosen equal to one) by the relation
\begin{eqnarray}\label{Eq20} F^{D^+}_{d\bar d\to S^0\to\eta \pi^0}(s)
=-\frac{1}{\sqrt{2}}F^{D^0}_{d \bar u\to S^-\to\eta \pi^-}(s)
\end{eqnarray} which follows from the assumption of equality of the
$a_0(980)^0$ and $a_0(980)^-$ masses and isotopic symmetry for the
coupling constants. It is easy to check using the following
relations
\begin{eqnarray}\label{Eq21} g_{d\bar d\eta\pi^0}=-g_0\sin
(\theta_i-\theta_p),\ \ \ g_{d\bar d\eta'\pi^-}=-g_0\cos
(\theta_i-\theta_p),\ \ \ g_{d\bar dK^0 \bar K0}=g_0,\qquad\qquad\quad\\
\label{Eq22} g_{a^0_0\eta\pi^0}=\bar{g}\cos(\theta_i- \theta_p),\ \
\ g_{a^0_0\eta'\pi^0}=-\bar{g}\sin( \theta_i- \theta_p),\ \ \
g_{a^0_0K^+K^-}=\bar{g}/\sqrt{2},\ \ \ g_{a^0_0K^0 K^0}=-\bar{g}/
\sqrt{2},\end{eqnarray} together with Eqs. (\ref{Eq10}) and
(\ref{Eq14}). Relations in (\ref{Eq22}) take into account that the
flavor structure of the $a_0(980)^0$ wave function has the form
\cite{Ja77}
\begin{eqnarray}\label{Eq22a} a_0(980)^0=\frac{1}{
\sqrt{2}}(u\bar u-d\bar d)s\bar s=\frac{1}{
\sqrt{2}}\,\eta_s\pi^--\frac{1}{{2}}\,K^+K^-+\frac{1}{{2}}\,K^0\bar
K^0.\end{eqnarray} Of course, for $m_{a_0^-}=m_{a_0^0}$ Eq.
(\ref{Eq20}) may be slightly violated in the region of the $K\bar K$
thresholds, but the relation for the total decay widths
\begin{eqnarray}\label{Eq9}\frac{\Gamma(D^0\to\eta\pi^-
e^+\nu_e)}{\Gamma(D^+\to\eta\pi^0 e^+\nu_e)}=2\end{eqnarray} should
hold very well. However, it should be noted here that the shapes of
the $\eta\pi^-$ and $\eta\pi^0$ mass spectra near the $K\bar K$
thresholds are very sensitive to the $a_0(980)$ mass and this fact
can be used to estimate the mass difference of the $a_0(980)^-$ and
$a_0(980)^0$ states in the $D$ decays. For example, if
$m_{a_0^0}\approx0.985$ GeV and $m_{a_0^-} \approx 0.995$ GeV
\cite{PDG2020,BESIII17,Te98,Ac17,AK18a}, then this reduces the ratio
(\ref{Eq9}) by $(5-6)$\%.

\section{\boldmath $a_0(980)$ resonance in the $D\to K\bar Ke^+\nu_e$ decays}

Production of the subthreshold $a_0(980)^-$ resonance in the decay
$D^0\to K^0K^-e^+\nu_e$ is strongly (at least an order of magnitude)
suppressed by the phase space in comparison with its production in
$D^0\to\eta\pi^-e^+\nu_e$. Certainly, experimental investigations of
the $D\to K\bar Ke^+\nu_e$ decays near the $K\bar K$ thresholds is a
difficult problem. A detailed theoretical analysis of these decays
will be urgent as soon as it becomes possible to carry out the
corresponding measurements. Therefore, here we only briefly describe
the characteristic features of the $K\bar K$ mass spectra associated
with the manifestation of the $a_0(980)$ resonance in our model.

\begin{figure} 
\begin{center}\includegraphics[width=7.5cm]{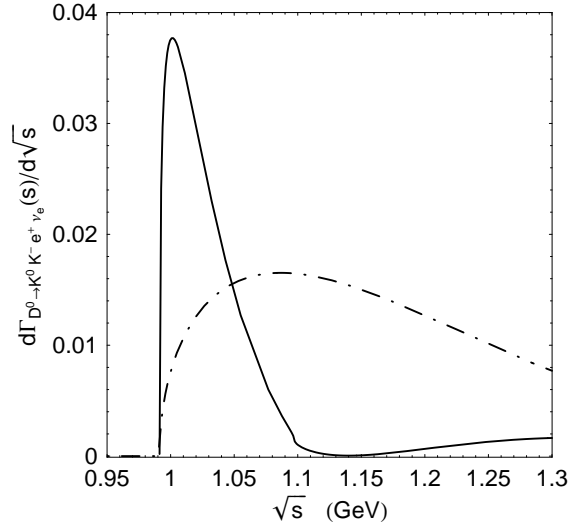}
\caption{\label{Fig9} The solid curve shows manifestation of the
$a_0(980)^-$ resonance in the $K^0K^-$ mass spectrum in the decay
$D^0\to K^0K^- e^+\nu_e$. The dash-dotted curve shows the
contribution of the point-like $K^0K^-$ production amplitude [i.e.,
$g_0$ in Eq. (\ref{Eq23})]. The scale along the ordinate axis is the
same as in Fig. \ref{Fig7}.}\end{center}\end{figure}

By analogy with Eq. (\ref{Eq18}), we write the amplitude for the
$D^0\to K^0K^-e^+\nu_e$ decay in the form
\begin{eqnarray}\label{Eq23} F^{D^0}_{d\bar u\to S^-\to K^0K^-}(s)=g_0
\left[1+I_{K^0 K^-}(s)\frac{\bar{g}^2}{16\pi D_{a^-_0}(s)}\right]+
g_0\sqrt{2}\sin\vartheta\cos\vartheta\left(I_{\eta \pi^-}(s)-I_{
\eta'\pi^-}(s)\right)\frac{\bar{g}^2}{16\pi D_{a^-_0}(s)}.
\end{eqnarray}
The corresponding $K^0K^-$ mass spectrum is shown in Fig.
\ref{Fig9}. A feature of this spectrum is the strong destructive
interference between the amplitude of the point-like production of
$K^0K^-$ [i.e., $g_0$ in Eq. (\ref{Eq23})] and the amplitudes
containing the contribution from the $a_0(980)^-$ resonance. As a
result, there arises a resonancelike structure in the region
$m_{K^0}+m_{K^-}<\sqrt{s}< 1.1$ GeV.

\begin{figure} 
\begin{center}\includegraphics[width=7.5cm]{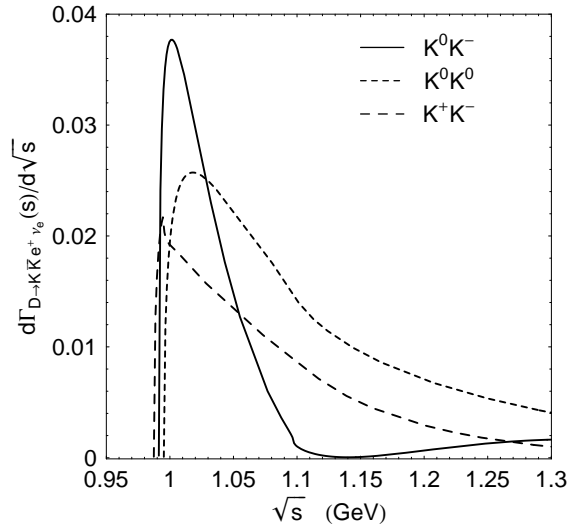}
\caption{\label{Fig10} The $K\bar K$ mass spectra in the decays
$D\to K\bar Ke^+ \nu_e$ caused by the $a_0(980)$ resonance
production. The scale along the ordinate axis is the same as in Fig.
\ref{Fig7}.} \end{center}\end{figure}

The $a_0(980)^0$ contributions to the amplitudes of the $D^+\to
K^0\bar K^0e^+\nu_e$ and $D^+\to K^+K^-e^+\nu_e$ decays are
\begin{eqnarray}\label{Eq24} F^{D^+}_{d\bar d\to S^0\to K^0\bar K^0}
(s)=g_0\left[1+I_{K^0\bar K^0}(s)\frac{\bar{g}^2}{32\pi D_{a^0_0}(s
)}\right]+ g_0\sin\vartheta\cos\vartheta\left(I_{\eta\pi^-}(s)-I_{
\eta'\pi^-} (s)\right)\frac{\bar{g}^2 }{\sqrt{2}16\pi D_{a^0_0}(s)},
\end{eqnarray}
\begin{eqnarray}\label{Eq25} F^{D^+}_{d\bar d\to S^0\to K^+K^-}(s)=-g_0
I_{K^+K^-}(s)\frac{\bar{g}^2}{32\pi D_{a^0_0}(s)}-g_0\sin\vartheta
\cos\vartheta\left(I_{\eta \pi^-}(s)-I_{\eta'\pi^-}(s)\right)\frac{
\bar{g}^2}{\sqrt{2}16\pi D_{a^0_0}(s)}.\end{eqnarray} The
corresponding $K^0\bar K^0$ and $K^+K^-$ mass spectra are shown in
Fig. \ref{Fig10}, together with the $K^0K^-$ one. Here we pay
attention to the dominance of the decay channel into $K^0\bar K^0$
over the $K^+K^-$ channel.
In fact, the decays $D^+\to K^0\bar K^0e^+\nu_e$ and $D^+\to
K^+K^-e^+\nu_e$ are more complicated than the decay $D^0\to
K^0K^-e^+\nu_e$, as they can also contain the contribution from the
isoscalar $f_0(980)$ resonance.

\section{\boldmath Conclusion}

A simple model of the four-quark mechanism of the $a_0(980)$
resonance production in the decays $D^0\to\eta\pi^-e^+\nu_e$ and
$D^+\to\eta\pi^0e^+\nu_e$ is constructed. It is shown that the
characteristic features of the shape of the $\eta\pi^-$ and
$\eta\pi^0$ mass spectra can serve as the indicator of the
production mechanism and internal structure of the $a_0(980)$ state.
Future experiments with high statistics on the decays $D\to
a_0(980)e^+\nu_e\to\eta\pi e^+\nu_e$ are highly demanded by the
physics of light scalar mesons.\\[0.5cm]

\begin{center} {\bf ACKNOWLEDGMENTS} \end{center}

The work was carried out within the framework of the state contract
of the Sobolev Institute of Mathematics, Project No.
0314-2019-0021.\\

\begin{center} {\bf APPENDIX:\, THE FUNCTION {\boldmath
$\widetilde{I}_{ab}(s)$}} \end{center}


\setcounter{equation}{0}
\renewcommand{\theequation}{A\arabic{equation}}

The dispersion integral $\widetilde{I}_{ab}(s)$ defined in Eq.
\ref{Eq11} is given by: for $s>m_{ab}^{(+)\,2}$
\begin{eqnarray}\label{A1} \widetilde{I}_{ab}(s)=\frac{s}{\pi}
\int\limits^\infty_{m_{ab}^{(+)\,2}}\frac{\rho_{ab}(s')\,ds'}{
\,s'(s'-s-i\varepsilon)}=L_{ab}(s)+\rho_{ab}(s)
\left(i-\frac{1}{\pi}\,\ln\frac{\sqrt{s-m_{ab}^{(-)
\,2}}+\sqrt{s-m_{ab}^{(+)\,2}}}{\sqrt{s-m_{ab}^{(-)\,2}}-\sqrt{s
-m_{ab}^{(+)\,2}}}\right),
\end{eqnarray} where
$m_{ab}^{(\pm)}=m_a\pm m_b$,\, $m_a>m_b$,\, $\rho_{ab}(s)=
\sqrt{s-m_{ab}^{(+)\,2}} \,\sqrt{s-m_{ab}^{(-)\,2}}\,/s$,\,  and
$$L_{ab}(s)=\frac{1}{\pi}\left[1+\left(\frac{m_{ab}^{(+)\,2}+
m_{ab}^{(-)\,2}}{2m_{ab}^{(+)}m_{ab}^{(-)}}-\frac{m_{ab}^{(+)}
m_{ab}^{(-)}}{s}\right)\ln\frac{m_a}{m_b}\right];$$ for
$m_{ab}^{(-)\,2}<s<m_{ab}^{(+)\,2}$
\begin{eqnarray}\label{A2}
\widetilde{I}_{ab}(s)=L_{ab}(s)-\rho_{ab}(s)\left(1-\frac{2}{\pi}
\arctan\frac{\sqrt{ m_{ab}^{(+)\,2}-s}}{\sqrt{s-m_{ab}^{
(-)\,2}}}\right),
\end{eqnarray} where  $\rho_{ab}(s)$\,=\,$\sqrt{m_{ab}^{(+)\,2}-s}
\,\sqrt{s-m_{ab}^{(-)\,2}}\,/s$;\, and for $s<m_{ab}^{(-)\,2}$
\begin{eqnarray}\label{A3}\widetilde{I}_{ab}(s)=L_{ab}(s)+\frac{\rho_{ab}(s)
}{\pi}\,\ln\frac{ \sqrt{m_{ab}^{(+)\,2}-s}+\sqrt{m_{ab}^{(-)\,2
}-s}}{\sqrt{m_{ab}^{(+)\,2}-s}-\sqrt{m_{ab}^{(-)\,2}-s}},
\end{eqnarray}
where $\rho_{ab}(s)$\,=\,$\sqrt{m_{ab}^{(+)\,2}-s}\,\sqrt{
m_{ab}^{(-)\,2}-s}\,/s$. If $b=\bar a$,  then $m_b=m_a$ and
$\widetilde{I}_{ab}(s)=\widetilde{I}_{a\bar a}(s)$ has a more simple
form:
\begin{eqnarray}\label{A4} \widetilde{I}_{a\bar a}(s)& = &\frac{2}{\pi}+\rho_{a\bar a}(s)
\left(i-\frac{1}{\pi}\,\ln\frac{1+\rho_{a\bar a}(s)}{1-\rho_{a\bar
a}(s)}\right),\quad\mbox{for }s>4m^2_a,\ \ \mbox{  and}\\ \label{A5}
\widetilde{I}_{a\bar a}(s)& = &\frac{2}{\pi}-|\rho_{a\bar
a}(s)|\left(1-\frac{2}{\pi}\,\arctan|\rho_{a\bar a}(s)|\right),
\quad\mbox{for }0<s<4m^2_a,
\end{eqnarray} where $\rho_{a\bar a}(s)=\sqrt{1-4m^2_a/s}\,$.



\begin{thebibliography}{99}
\bibitem{CLEO09b}  K. M. Ecklund {\it et al.} (CLEO Collaboration), Phys. Rev. D {\bf 80}, 052009 (2009).    
\bibitem{WL10} W. Wang and C. D. L\"{u}, Phys. Rev. D {\bf 82}, 034016 (2010).                               
\bibitem{Pe10} M. R. Pennington, AIP Conf. Proc. {\bf 1257}, 27 (2010).
\bibitem{Fa11} A. H. Fariborz, R. Jora, J. Schechter, and  M. N. Shahid, Phys. Rev. D {\bf 84}, 094024 (2011).  
\bibitem{AK12} N. N. Achasov and A. V. Kiselev, Phys. Rev. D {\bf 86}, 114010 (2012).                        
\bibitem{PDG2020} P. A. Zyla {\it et al.} (Particle Data Group), Prog. Theor. Exp. Phys. {\bf 2020}, 083C01 (2020).
\bibitem{BABAR08}  B. Aubert {\it et al.} ({\it BABAR} Collaboration), Phys. Rev. D {\bf 78}, 051101(R) (2008).  
\bibitem{CLEO09a}  J. Yelton {\it et al.} (CLEO Collaboration), Phys. Rev. D {\bf 80}, 052007 (2009).
\bibitem{BESIII18} M. Ablikim {\it et al.} (BESIII Collaboration), Phys. Rev. Lett. {\bf 121}, 081802 (2018). 
\bibitem{BESIII19} M. Ablikim {\it et al.} (BESIII Collaboration), Phys. Rev. Lett. {\bf 122}, 062001 (2019). 
\bibitem{BESIII21} M. Ablikim {\it et al.} (BESIII Collaboration), Phys. Rev. D {\bf 103}, 092004 (2021).     
\bibitem{AK14} N. N. Achasov and A. V. Kiselev, Int. J. Mod. Phys. Conf. Ser. {\bf 35}, 1460447 (2014).
\bibitem{SO15} T. Sekihara and E. Oset, Phys. Rev. D {\bf 92}, 054038 (2015).                           
\bibitem{Sh17} Y. J. Shi, W. Wang, and S. Zhao, Eur. Phys. J. C {\bf 77}, 452 (2017).                   %
\bibitem{Ch17} X. D. Cheng, H. B. Li, B. Wei, Y. G. Xu, and M. Z. Yang, Phys. Rev. D {\bf 96}, 033002 (2017).
\bibitem{AK18} N. N. Achasov and A. V. Kiselev, Phys. Rev. D {\bf 98}, 096009 (2018).                   
\bibitem{AK19} N. N. Achasov and A. V. Kiselev, EPJ Web Conf. {\bf 212}, 03001 (2019).
\bibitem{Ac20} N. N. Achasov, Phys. Part. Nucl., {\bf 51}, 632 (2020).
\bibitem{So20} N. R. Soni, A. N. Gadaria, J. J. Patel, and J. N. Pandya, Phys. Rev. D {\bf 102}, 016013 (2020). 
\bibitem{AKS20}N. N. Achasov, A. V. Kiselev, and G. N. Shestakov, Phys. Rev. D {\bf 102}, 016022 (2020).        
\bibitem{Sh21a}Y. J. Shi, C. Y. Seng, F. K. Guo, B. Kubis, U.-G. Mei{\ss}ner, and W. Wang, J. High Energy Phys. {\bf 04} (2021) 086. 
\bibitem{Sh21b}Y. J. Shi and U.-G. Mei{\ss}ner, Eur. Phys. J. C {\bf 81}, 412 (2021).                           
\bibitem{Ja77} R. L. Jaffe, Phys. Rev. D {\bf 15}, 267 (1977); {\bf 15}, 281 (1977).
\bibitem{AI89} N. N. Achasov and V. N. Ivanchenko, Nucl. Phys. {\bf B315}, 465 (1989).
\bibitem{Ac21} N. N. Achasov, J. V. Bennett, A. V. Kiselev, E. A. Kozyrev, and G. N. Shestakov, Phys. Rev. D {\bf 103}, 014010 (2021).
\bibitem{Ab15} M. Ablikim {\it et al.} (BESIII Collaboration), Phys. Rev. D {\bf 92}, 052003 (2015).             
\bibitem{Ba86} R. M. Baltrusaitis {\it et al.} (MARK III Collaboration), Phys. Rev. Lett. {\bf 56}, 2140 (1986). 
\bibitem{Sa11} P. del Amo Sanchez {\it et al.} ({\it BABAR} Collaboration), Phys. Rev. D {\bf 83}, 072001 (2011).
\bibitem{Ok63} S. Okubo, Phys. Lett. {\bf 5}, 165 (1963).
\bibitem{Ok78} S. Okubo, Progr. Theor. Phys. {\bf 63}, 1 (1978).
\bibitem{Zw80} G. Zweig, in {\it Developments in the Quark Theory of Hadrons}, edited by D.B. Lichtenberg and S.P.
               Rosen (Hadronic Press, Massachusetts, 1980).
\bibitem{Ii66} J. Iizuka, Prog. Theor. Phys. Suppl. {\bf 37}, 21 (1966).
\bibitem{Lip87}H. J. Lipkin, Nucl. Phys. {\bf B291}, 720 (1987).
\bibitem{Lip96}H. J. Lipkin and B. S. Zou, Phys. Rev. D {\bf 53}, 6693 (1996).
\bibitem{Fl76} S. M. Flatte, Phys. Lett. {\bf 63B}, 224 (1976).
\bibitem{ADS80}N. N. Achasov, S. A. Devyanin, and G. N. Shestakov, Phys. Lett. {\bf 96B}, 168 (1980).
\bibitem{BESIII17}M. Ablikim {\it et al.} (BESIII Collaboration), Phys. Rev. D {\bf 95}, 032002 (2017).
\bibitem{Wa52} K. M. Watson, Phys. Rev. {\bf 88}, 1163 (1952).
\bibitem{Om58} R. Omn\`{e}s, Nuovo Cimento A {\bf 8}, 316 (1958).
\bibitem{GRS69}M. Gourdin, F.M. Renard, and L. Stodolsky, Phys. Lett. {\bf 30B}, 347 (1969).
\bibitem{Te98} S. Teige {\it et al.} (E852 Collaboration), Phys. Rev. D {\bf 59}, 012001 (1998).
\bibitem{Ac17} S. Acharya {\it et al.} (ALICE Collaboration), Phys. Lett. B 774, 64 (2017).
\bibitem{AK18a}N. N. Achasov and A. V. Kiselev, Phys. Rev. D {\bf 97}, 036015 (2018).
\end{thebibliography}
\end{document}